\def\year{2016}\relax
\pdfoutput=1
\documentclass[letterpaper]{article}
\usepackage{aaai16}
\usepackage{times}
\usepackage{helvet}
\usepackage{courier}
\usepackage{graphicx} 
\usepackage{subfigure}
\usepackage[table]{xcolor}

\newcommand\tabhead[1]{\small\textbf{#1}}

\begin{document}
\title{Networks of Gratitude: Structures of Thanks and User Expectations in Workplace Appreciation Systems}
\author{Emma S. Spiro\\
University of Washington\\
Mary Gates Hall, Suite 370\\
Seattle, WA 98195\\
\And
J. Nathan Matias\\
MIT Media Lab\\
75 Amherst St\\
Cambridge, MA 02139\\
\And
Andr\'{e}s Monroy-Hern\'{a}ndez\\
Microsoft Research \\
One Microsoft Way \\
Redmond, WA 98052\\
}
\maketitle
\begin{abstract}
Appreciation systems\textemdash platforms for users to exchange thanks and praise\textemdash are becoming common in the workplace, where employees share appreciation, managers are notified, and aggregate scores are sometimes made visible. Who do people thank on these systems, and what do they expect from each other and their managers? After introducing the design affordances of 13 appreciation systems, we discuss a system we call \emph{Gratia}, in use at a large multinational company for over four years. Using logs of 422,000 appreciation messages and user surveys, we explore the social dynamics of use and ask if use of the system addresses the recognition problem. We find that while thanks is mostly exchanged among employees at the same level and different parts of the company, addressing the recognition problem, managers do not always act on that recognition in ways that employees expect.
\end{abstract}

\noindent Appreciation systems \textemdash a genre of messaging and microblogging systems that mediate digital expressions of appreciation in the workplace, have become widely adopted in recent years. 35\% of companies used some form of online peer recognition system in 2015, an approach that was overtaking top-down recognition efforts \cite{lahey_art_2015}. Since these systems track appreciation between workers and produce reports for managers, they also delegate parts of employee evaluation to a wider peer network. 

Examples of appreciation systems include KudosNow \cite{short_kudos_2011}, which mediates exchanges of peer thanks online, aggregating that appreciation to produce performance reports that support managerial decisions. On Yammer, an enterprise social networking system (SNS), employees can  ``praise" each other, incrementing a per-employee praise count \cite{yammer_yammer_2011}. The peer bonus system Bonus.ly gives employees a monthly budget to reward the colleagues who they appreciate \cite{naziri_bonus.ly_2013}. Other systems display interpersonal appreciation and group metrics on screens and artwork in physical workplaces \cite{yoon_thank_2013,munson_thanks_2011}. 

By design, the thanks, bonuses, or praise sent on these systems are personal exchanges that also reach wider audiences: the receiver's manager, the sender's manager, and sometimes others in the company (where profiles, reports or public displays are involved). While the recipient sees an individual message of appreciation, a manager might see recent appreciation sent or received by her team, and an executive might see aggregate reports of praise or thankfulness across teams. The design of these reports implies an expectation that they can support decisions about feedback, promotion, team composition, and team well-being.

In this paper, we identify two fundamental questions for understanding the use of any appreciation system in the workplace: (a) who do users thank, and (b) what outcomes do users expect from participation? 

By considering the structure of \textit{who users thank}, we can evaluate rhetoric from designers that appreciation systems mediate relationships across company hierarchies, as well as concerns that these systems merely offer ``democratized kissing up" \cite{jurgen_peer--peer_2015}. In one possibility, thanks might be primarily sent from managers to subordinates, reproducing hierarchical structures of feedback that the appreciation system was designed to supplement. Alternatively, employees could ``thank up'' to managers to signal indebtedness and perhaps gain favor or ``thank across'' the company hierarchy, either to their immediate peers or to others elsewhere in the organizational hierarchy. 

\textit{What do users expect} from participation? A thanks message may signal expectations from the parties exchanging thanks, as well as expectations of the managers who have access to reports on employee appreciation. For example, when thanking someone through an appreciation system, the sender may expect the receiver of thanks to feel appreciated while also expecting the receiver's manager to account for the message in an upcoming performance review. Manager involvement or even the anticipation of manager involvement may affect this interpersonal exchange. On the other hand, lack of manager attention is also cited as a major problem for these appreciation systems \cite{lahey_art_2015}.

To answer these questions, we situate appreciation systems within related research, then describe the genre of appreciation systems, their design rhetoric, and major affordances across 13 systems. We then contribute a case study of over 422,000 thanks records over four years in the \emph{Gratia} appreciation system, a system designed by employees of a large multinational company and widely adopted throughout the company. We answer questions on who users thank with descriptive statistics and network analysis of thanks between users, comparing its structure to the organizational hierarchy. We answer questions about what users expect with content analysis of system logs and surveys of appreciation senders, receivers, and managers. 

\section{Related Work}

\subsection{Thanks and Gratitude}

This study includes analysis of the structural qualities (i.e. social networks) of appreciation that are mediated by appreciation systems. Theories on the structure and nature of thanks offer frameworks for studying behavior on these systems. Expressions of thanks are a signal of informal exchange between two parties that acknowledges a kind of social indebtedness distinct from formal trade or contracts \cite{smith_theory_1759,simmel_gratitude_1950,mccullough_is_2001}. These messages signal that two parties ``have now completed a (usually pleasing) reciprocal exchange, and the door is opened to the possibility of new and mutually pleasing exchanges in the future'' \cite{emmons_gratitude_2004}. 

This condition of social and emotional indebtedness between parties is described by the Stoic philosopher Seneca: ``He who receives a benefit with gratitude, repays the first instalment of it'' \cite{seneca_benefeciis_1928}. The two-person structure of thanks has been observed in peer resource exchange systems, where users who are unable to generalize their sense of obligation to a wider community share tokens of appreciation with the specific parties who helped them to ``lessen uneasy feelings of indebtedness'' \cite{lampinen_indebtedness_2013}. Thanks is also usually attached to a single exchange between the parties, in contrast with gratitude, which is more often associated with overall life narratives, or attached to deities rather than people or institutions \cite{mccullough_is_2001}. 

\subsection{Workplace Social Network Systems}

Corporate SNS have broadened information seeking within companies \cite{ingebricson_corporate_2010}, expanding the ability of employees to find and connect with people outside their immediate teams \cite{dimicco_people_2009}, participating in a trend that emphasizes the value of informal social networks in contrast with strict organizational hierarchies \cite{krackhardt_informal_1993,burt_structural_2004}. In addition to facilitating networks of social discovery and support, these systems often expose network activity to managers who oversee employee goals, incentives, and evaluation within the organization.

Companies have been adopting blogging, microblogging, wikis, and SNS to facilitate knowledge sharing about people and ideas across teams. These systems act as a virtual ``water cooler'' promoting information exchange and relationship formation. In corporate microblogging systems, employees post personal updates, company news, questions, and requests for discussion \cite{zhang_case_2010}. Users of corporate blogging systems report benefits of communicating ``informally," ``without bureaucracy,'' and ``meeting people from other parts of the company/outside my circle,'' as they make posts and add comments. Blogging becomes an ``informal mechanism that links disparate, far-flung parts of the organization into constructive contact" \cite{jackson_corporate_2007}. 

Informal exchanges in the workplace and the networks they form, such as friendship or advice-seeking networks, are known to be important for understanding how work-related tasks are accomplished \cite{krackhardt_informal_1993,whittaker_informal_1994}. By identifying and understanding these ``invisible'' relationships, managers hope to gain insight on informal leadership, information flows, and structures of trust. Other research on informal networks within organizations has focused on identifying sources of good or novel ideas, comparing positions individuals may occupy in both the informal and formal structure, and assessing the capacity of informal networks to execute those ideas \cite{burt_structural_2004}. To identify and measure these networks, researchers have examined self-reported relationships \cite{krackhardt_informal_1993}, email communications \cite{donath_visual_1995,gilbert_phrases_2012} and physical proximity \cite{wu_mining_2008}. Appreciation represents another signal of these informal networks.

Workplace SNS are often used for relationship building and ``people sensemaking'' across organizational hierarchies as users discover and draw conclusions about other employees whom they do not already know \cite{dimicco_people_2009}. In these SNS, profile browsing activity supports the creation of new relationships beyond current teams (i.e. \textit{bridging} social capital) \cite{dimicco_people_2009}, while conversation and interaction strengthen existing relationships (i.e. \textit{bonding} social capital)\cite{putnam_bowling_2001,steinfield_bowling_2009}. The separation of these internal workplace systems from public SNS and microblogging platforms is important to many employees; in one longitudinal study, employees reported resistance and skepticism about the use of public SNS to share information among colleagues \cite{archambault_longitudinal_2012}.

\subsection{Recognition and Incentives in Informal Networks}

Systems for informal information sharing and relationship development may operate in tension with formal management structures, which are not always able to acknowledge or reward activity in informal networks. For example, several studies on the failed adoption of computer-supported cooperative work systems cite a lack of recognition or incentives as one major risk to system adoption \cite{hasan_wiki:_2006}. In one company with an ``up or out" promotion structure, where cooperation among employees was not included in promotion review, employees avoided information sharing software \cite{orlikowski_learning_1992}. Concerns that cooperation systems ``undermine management authority" can even lead to managerial resistance \cite{grudin_why_1988}. To address this tension, Grudin argues that systems should offer distinct functions for individual contributors, managers, and executives, in ways that are compatible with incentive structures and the social conventions of teams \cite{grudin_managerial_2004}.

The genre of appreciation systems addresses the incentives problem in informal cooperation by combining peer incentives with the hierarchical management structure, supporting all three of Grudin's suggested audiences for successful workplace cooperation systems. When an \emph{individual contributor} sends or receives appreciation, they support each other through the informal relational network. At the same time, their \emph{managers} can see the appreciation and incorporate that information into review and promotion processes. Appreciation systems that generate team-level reports offer summaries that can be skimmed by \emph{executives}, who cannot review each appreciation across all their teams. Since appreciation systems make thanks visible within informal and formal structures, they have possible formal outcomes beyond the informal relationships between people as managers and executives view these reports. 

Despite this potential, the use of an appreciation systems in a specific workplace only addresses the recognition problem if two conditions are met: (a) if the structure of appreciation represents exchanges that managers wouldn't see otherwise, and (b) if managers incorporate peer recognition into decisions affecting their teas. By asking who people thank and what people expect from their use of appreciation systems, this paper observes the applicability of these systems to the recognition problem.

\section{Appreciation Systems}

\subsection{Defining Appreciation Systems}

To describe the genre of appreciation systems, we present an overview of publicly-documented systems. Our initial set of 55 potential systems was created by reviewing over 150 English language news articles and press releases, reviewing feature descriptions of corporate SNS, and asking designers about competitors.\footnote{Press releases were discovered through PR Newswire.} Systems were retained if they featured peer appreciation and also made appreciation visible to managers. Systems were excluded if they were not designed for the workplace (i.e. Wikipedia's peer thanks system), if appreciation was purely private (i.e. workplace e-card systems), if only managers could send appreciation (i.e. performance review systems), or if full public documentation of features was unavailable. Of the initial set, we focus on the 13 appreciation systems included in Table~\ref{tab:affordances}.

\rowcolors{2}{gray!10}{white}
\def\arraystretch{1.2}%
\begin{table*}[ht!]
\centering
\begin{scriptsize}
\begin{tabular}{|l|p{1.6cm}llp{1.2cm}p{1.4cm}p{1.2cm}p{1.4cm}ll|}
\hline
\rowcolor{gray!30}
\tabhead{System} & \tabhead{Audience} & \tabhead{Thanks} & \tabhead{Feedback} & \tabhead{Points/\newline Credit} & \tabhead{Rewards} &  \tabhead{Badges} & \tabhead{Reporting\newline Analytics} & \tabhead{Moderator} & 
\tabhead{Limits}\\
\hline
\emph{Gratia} & receiver, managers & yes & no & no &  no & no & team reports, API & no & 2/wk\\
Yammer & receiver, \newline profile page & yes & no & no & no & yes & API & no & no\\
KudosNow & receiver, manager & yes & yes & yes & online store & yes & worker and team reports & yes & yes\\
Bonus.ly & company-wide & no & no & yes & bonus & no & worker & yes & yes \\
Thanks Duck & Twitter, physical & yes & no & no & no & no &  workspace\newline display & no & no\\
SI Display & Twitter, public screen & yes & no & no & no & no & no & no & no \\
Braavoo & receiver, manager & no & yes &no & no & no & worker and team reports & no & no \\
7Geese & receiver, (opt)manager, (opt)team & no & yes & yes & no & yes & worker and team reports & no & no \\
HerdWisdom & receiver, manager & yes & yes & yes & yes & yes & worker and team reports & no & no \\
Work.com & receiver, \newline profile page & yes & yes & yes & no & yes & team reports, API& yes & yes \\
GiveAWow & receiver, manager & no & yes & yes & yes & yes & worker and team reports, API & yes & no \\
O.C. Tanner & receiver, wall of fame & yes & yes & optional & optional & digital and physical & worker and team reports, API & unknown & unknown \\
Gthanks & receiver, \newline profile page & yes & yes & yes & bonus & yes & unknown & yes & no \\
\hline
\end{tabular}
\end{scriptsize}
\caption{Affordances of workplace appreciation systems}
\label{tab:affordances}
\end{table*}

Our term ``\textit{appreciation system}'' is chosen to reflect these inclusion criteria and connect these systems to scholarship on the exchange of thanks. We also considered the term ``gratitude systems,'' but this term is inconsistent with theory \textemdash gratitude is an affect rarely attached to a specific exchange \cite{mccullough_is_2001}. Other terms like ``thanks systems" and ``peer incentive systems'' over-emphasize the peer aspect of these systems or over-emphasizes a single theory of motivation. ``Appreciation systems,'' is fruitfully ambiguous. It can include both interpersonal appreciation and formal appreciation from one's manager in a wide range of forms: a private note, a bonus, or more public praise.

\subsection{Technical Affordances of Appreciation Systems}
We present in Table~\ref{tab:affordances} an overview of the technical affordances of appreciation systems. While all systems in the table make appreciation visible to the receiver and to managers, they still vary widely in the \textit{audience} and visibility of appreciation. In some cases, appreciation is associated with a worker's SNS profile page, while other systems place individual appreciation on a ``wall of fame" or ``Thank You Board" \cite{munson_thanks_2011}. Physical display systems sometimes show a team's aggregated appreciation rather than individual thanks \cite{yoon_thank_2013}.

Not all appreciation systems support interpersonal \textit{thanks}, which is described as distinct from \textit{feedback}. While feedback can often be included in the text of an appreciation system, many systems allow managers to structure participation by requiring appreciation to include ratings in the categories used for worker performance reviews. Descriptions of \textit{feedback} affordances emphasize appreciation given for work-related activities within a formal performance review system. In contrast, systems that promote \textit{thanks} encourage users to share appreciation for anything that might prompt thanks toward the recipient. In some systems, peer thanks and peer feedback affordances are distinct from each other.

Beyond feedback, another way to quantify appreciation is to award \textit{points or credit} to receivers. In many cases, these points are exchanged for \textit{rewards} on an online store, associated with bonuses, or accumulated towards charitable donations. In other systems, appreciation points are associated with \textit{badges} or other forms of quantified reputation on a corporate SNS. \textit{Reporting and analytics} systems vary widely across appreciation systems, depending on the details of thanks, feedback, credit, reward, and badges. Most of these systems offer aggregated worker and team level reports, as well as API access.

Many appreciation systems offer affordances for \textit{moderation}, which allow managers or other moderators to approve or remove messages and bonuses. The designers of Google's GThanks peer bonus system initially avoided implementing moderation affordances, on a matter of principle of trust for employees \cite{bock_work_2015}. In addition to moderation, many systems, whether or not they offer rewards, include \textit{limits} and budgets on the thanks or credits that any worker can send.

\subsection{Design Rhetoric of Appreciation Systems}
In press releases and official blog posts, the design rhetoric for appreciation systems often focuses on highlighting unacknowledged contributions among peers. In promotional material, Bonus.ly argues that ``peers often reward an activity that would have gone completely unnoticed by managers" \cite{quinn_look_????}. In their FAQ page, the designers of \emph{Gratia} explain that ``recognition of achievements (especially when cross-group) doesn't have the right level of visibility,'' a problem that \emph{Gratia} is intended to address. Well-being and team bonding are also common design rhetoric for appreciation systems. Designer Ilwon Yoon, creator of the Thanks Duck, argues that appreciation is a deep human need, and that the duck is intended to ``create [a] more collaborative and cooperative environment" \cite{yoon_thank_2013}. 

\subsection{Top-Down, Informal, and Piggyback Deployments}
A wide range of workplace appreciation systems have been introduced in recent years through top-down and employee-initiated deployments, as features within existing corporate groupware, standalone systems, or ``piggyback" deployments. Among top-down systems, Yammer added a ``praise'' feature in 2011, inviting users to choose a ``praise badge,'' enter a message, and share their appreciation for individuals and groups \cite{yammer_yammer_2011}. In the peer bonus system Bonus.ly and Google's GThanks, employees are allocated budgets to reward other employees with praise and monetary bonuses. Depending on corporate policy, receivers may be paid directly, redeem bonuses in an online store, or in some cases, choose which charity will receive their bonus \cite{naziri_bonus.ly_2013,quinn_look_????,bock_work_2015}. 

Although these systems are typically introduced by managers or HR departments, appreciation systems have also been developed and introduced informally by employees. The \emph{Gratia} system we examine in this paper is a website-based appreciated system that was developed internally by a group of employees who wished to foster gratitude in the company. This system, which we describe in greater detail later, shares thanks messages among employees via email, which are shown to managers in digest form. 

\section{Case Study: The `\textit{\textbf{Gratia}}' Appreciation System}

In this paper, we examine the use of `\emph{Gratia}', a workplace appreciation system. \emph{Gratia} was developed as a grassroots effort by employees of a large multinational corporation who wanted to foster a culture of greater gratitude and appreciation within their company.\footnote{As requested by the company, all content and names, including the company's, have been anonymized. Also, although the system has always been internal, `\emph{Gratia}' is a pseudonym.} The system was later adopted by the I.T. department of the company and embedded into other systems, such as the internal people search engine. 

The company has a well-connected and geographically distributed workforce of nearly 120,000 people that actively uses a myriad of communication technologies: from email, to instant messaging, to video conferencing, to an internal social network site. Despite all those communication channels, `\emph{Gratia}' provides a single-purpose channel for thanks and has maintained an active user base. The company's workforce is clustered in a handful of large hierarchical organizations. These organizations were historically independent from one another, but in recent years have undergone a cultural shift towards more interdependence and collaboration. The organizations that make up the company have teams across multiple building and campuses around the world. Given the nature of `\emph{Gratia}', it is worth noting that typically employee yearly reviews are performed around June, while bonuses and promotions are distributed around August.

In the subsequent sections we explore the `\emph{Gratia}''s user experience, system usage, and social network of appreciation, with an analysis of system logs and user surveys. We focus our work on the structure of who users thank and what they expect from their participation. 

\begin{figure*}[ht!]
\subfigure[Sending thanks with \emph{Gratia}.]{
\includegraphics[height=1.45in]{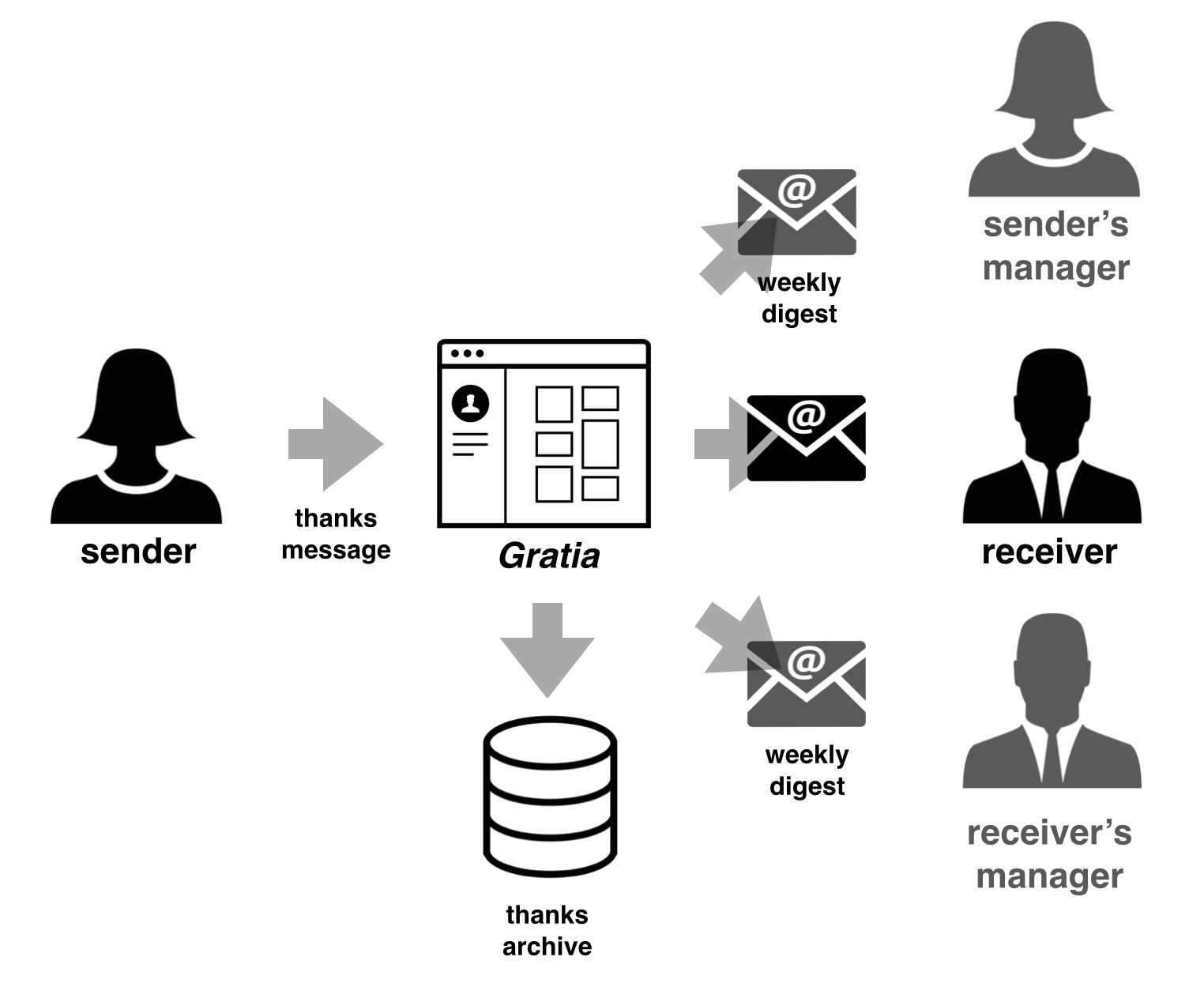}}
\subfigure[Anonymized mock-up of the \emph{Gratia} user interface.]{
\includegraphics[height=1.45in]{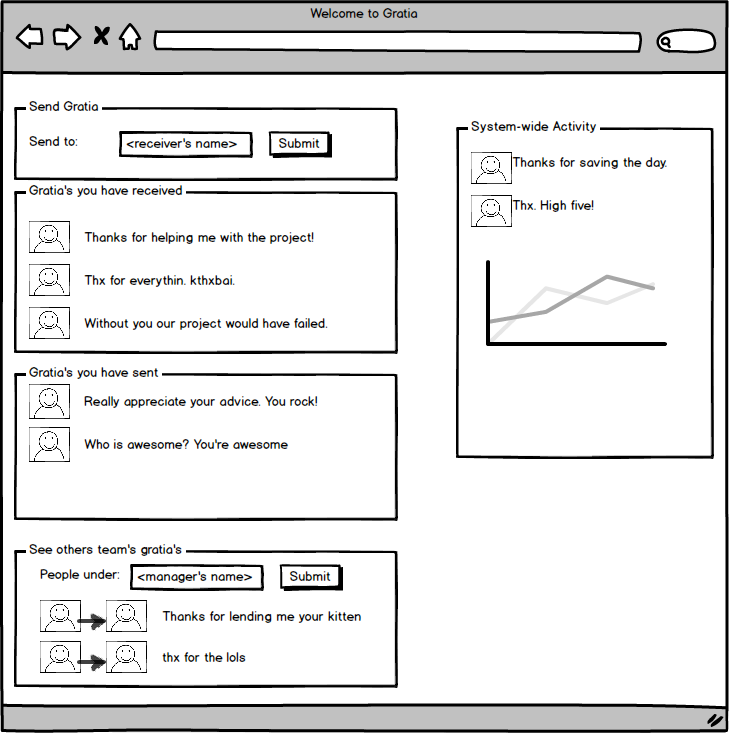}}
\subfigure[\emph{Gratia} user flow.]{
\includegraphics[height=1.45in]{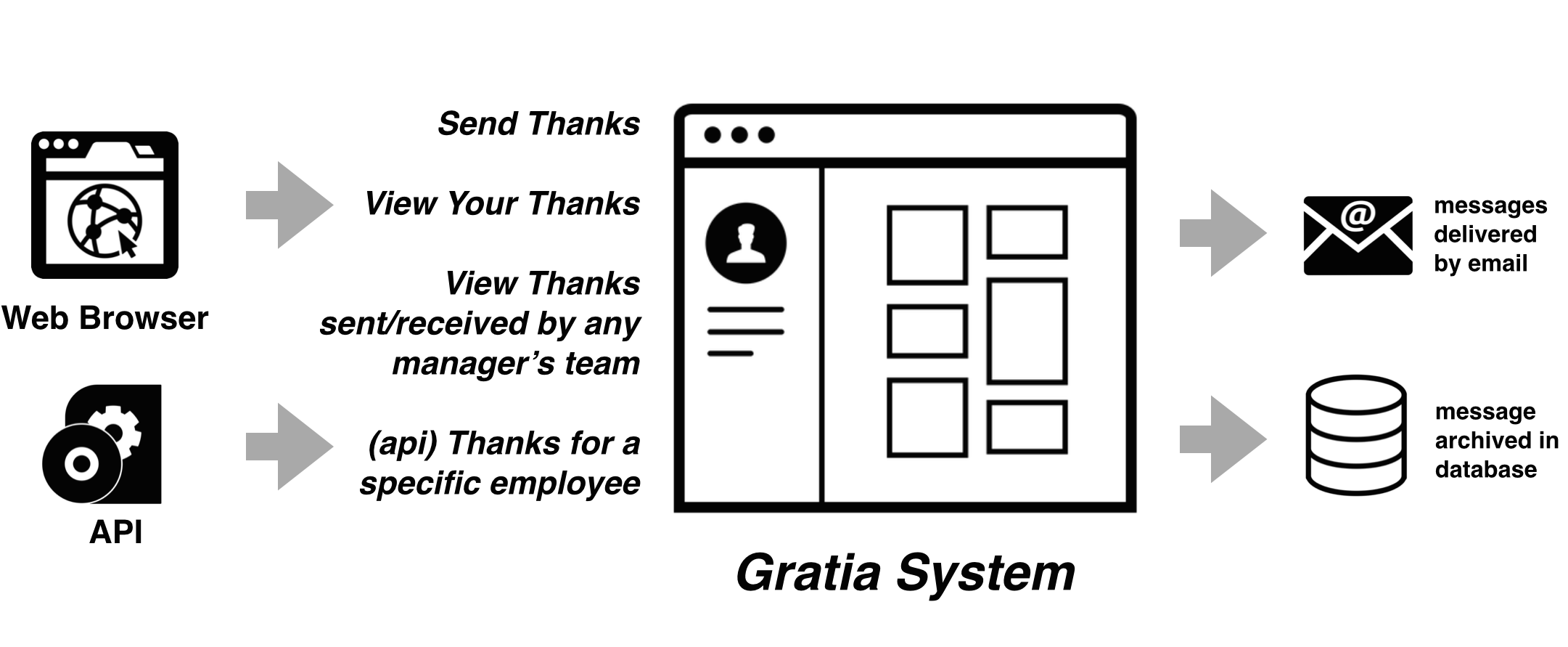}}
\caption{\emph{Gratia} system and features.}
\label{fig:sendingthanks}
\end{figure*}

\subsection{System Overview} 

To send thanks, employees visit the \emph{Gratia} web application and create a thanks message that is sent to the receiver via email. Company managers receive weekly digest emails describing all appreciation sent and received by their teams, as illustrated by Figure~\ref{fig:sendingthanks}(a). Using the \emph{Gratia} website, as seen in Figure~\ref{fig:sendingthanks}(b), anyone in the company can enter the name of a manager and see the thanks sent and received by that manager's team. Per-employee information is accessible via an Application Programming Interface (API) which has been integrated with other systems in the company, such as a contact directory, a mobile app, and a manager reporting system. The user flow of the system is described in Figure~\ref{fig:sendingthanks}(c).

In \emph{Gratia}, each employee is allocated a budget of two thanks per week. Thanks in \emph{Gratia} have no formal company-wide association with bonuses or other reward mechanisms, although managers may incorporate information from thanks messages into employee reviews. In internal communications, \emph{Gratia} is described as a ``peer recognition system" to help employees ``recognize helpful peers across the company.'' These communications explain who receives notification emails, what messages are visible across the company, and how thanks are aggregated for individuals and their managers.

\subsection{Data and Methods}

As part of our investigation of \emph{Gratia}, we analyze usage logs \textemdash approximately 422,000 records representing all thanks messages sent via the system over the four years from May 2010 to March 2014. These records include information identifying the sender and recipient of the thanks message along with the message text. Each record is timestamped. We combine \emph{Gratia} system log data with metadata about receivers and senders, that includes department name, job title, and manager. Managers are included in the data if at least one of their subordinates sent or received a thanks message during the observation period. While \emph{Gratia} usage logs represent longitudinal observations of thanks, employee records are sampled at the end of the observation period.

In our analysis we take a descriptive and relational approach to address: (1) who do users thank, and (2) what do users expect from participation? To address the first, our analysis focuses on features of the informal social network revealed through the act of sending a thanks message to a co-worker using the \emph{Gratia} system. \emph{Gratia's} network has just over 80,000 nodes (senders and receivers) and approximately 422,000 ties (thanks messages)\footnote{Statistics are approximate to help maintain confidentiality.}. Each node represents an employee; a directed tie from node $A$ to node $B$ exists when user $A$ sends a thanks message to user $B$. 

Managerial data from employee records allows us to construct a second network capturing the formal, organizational hierarchy \textemdash who manages whom. This organizational network takes on a strict tree structure; each employee has one and only one manager.\footnote{Recall: this tree represents a snapshot of the management hierarchy in March 2014 and does not capture promotions or management changes during this four year period. We took a follow-up sample in Dec 2014 to estimate the rate of job mobility and do not believe rates were high enough to be of concern for our analysis.}

To explore user expectations, we administered via email a short survey\footnote{the questions in the survey instrument are available at \texttt{http://bit.ly/1mPm2Wc}} to \emph{Gratia} users. We construct a stratified random sample based on interaction types, defining the sample strata by whether the participant has exclusively sent thanks messages, exclusively received thanks messages, or has sent and received thanks. We separately administer a survey to managers whose team members had interacted with \emph{Gratia} in any form (senders and/or receivers). 

Our sampling strategy was designed to ensure participants had diverse interactions with the \emph{Gratia} system, both in terms of the sender-receiver experience, as well as in their tenure on the system. We recruited participants who might have used the system a few years prior to the survey, as well as some who might have used Gratia only a few days prior. Although it is possible that participants' memories regarding the system might fade over time, the system is well-known in the company that at least awareness and opinions about were available. It is possible participants' opinions regarding the system might have changed over time, though, but we considered diversity of tenure to be more important to get a sense of the views about the system across time.

We obtained 29 responses from thanks senders,\footnote{Includes people who have exclusively sent thanks, as well as some that have both sent and received them} 44 responses from thanks recipients,\footnote{Includes people who have exclusively received thanks, as well as some that have sent and received them} and 15 responses from managers; yielding a response rate of 11\%, 17\% and 0.6\% respectively. Response rates were low, but we believe responses provide valuable qualitative insight for this study. Low response could be due to turnover in employees since we did not restrict the sample to recently active employees.

User surveys were designed to gain additional insight into user motivations and concerns. The questions were mostly multiple choice, with a few open-ended questions coded manually by the researchers. Specific questions were asked about how users first heard of \emph{Gratia}, about users' understanding of the general operation of the system, as well as about user reactions to incorporating thanks into formal performance reviews. Survey participants were also asked to reflect on preliminary findings (discussed in detail in subsequent sections) on the social norms of the system. Survey responses, which are presented in later sections, were analyzed by the authors to gain an understanding of common patterns and themes.

While some survey methodologies seek answers on individuals' experiences, our approach explicitly focused on how participants imagine their experience in context of the wider use of Gratia, which is why we also asked users these more generalized questions, such as why people higher or lower in the organizational behave in certain way, rather than asking the survey respondent to answer only about their experience.  

We begin our analysis with a short description of adoption and usage patterns over time to gain a basic understanding of the system itself; we go on to quantify structural features of the informal network of thanks and its relation to the formal company hierarchy to address the question of who do users thank. Finally, we incorporate qualitative data from the user survey to support interpretations and implications of the results. We make use of the \texttt{sna} package of the \textbf{R} statistical computing platform and Dato GraphLab Create\footnote{http://dato.com/products/create/overview.html} in our analysis.

\section{Findings}

\subsection{Messages on \textit{\textbf{Gratia}}}

Although messages on \emph{Gratia} are shared with both the intended recipient as well as managers, they typically employ a personal tone and context:\footnote{Examples have been slightly rephrased to preserve anonymity, but are consistent in tone and style to the original.}

\begin{itemize}
\item{``Thank you for the help you provided over the past few months in answering questions in easy understandable way. You are the best communicator in the team''}
\item{``Thanks Carl, You INSPIRE me! I wish the best for you, as you always said, I'm just an IM away!''}
\item{``Great job with rocking the project since you been here, don't get discouraged with the things you cannot control. Stay motivated!''}
\item{``Thanks for all the great work on the FooBar. I learned a lot working from you on it!''}
\item{``Alex, thank you for all your help in getting the foobar created. You went above and beyond to help.''}
\end{itemize}

\begin{figure}[ht!]
\includegraphics[width=\linewidth]{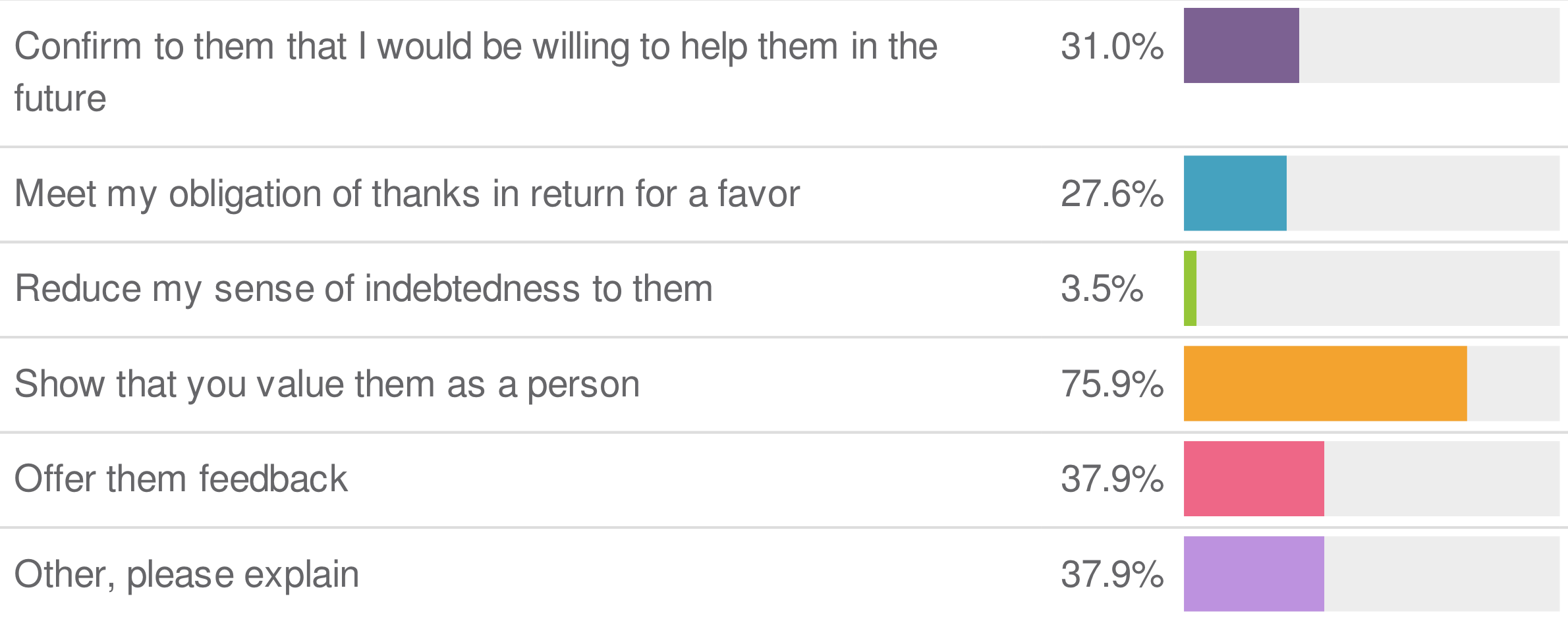}
\caption{User survey responses when asked about what they hoped to achieve with sending appreciation.\footnotemark}
\label{fig:kudos-achieve}
\end{figure}

The topic and purpose of appreciation messages ranged widely across feedback, encouragement, social value, and reciprocity. Likewise, in surveys, receivers reported that their appreciation messages were mostly in relation to a specific event or task (93\%), with over half reporting a general sense of appreciation (59\%) and more than a quarter identifying appreciation for a specific favor (46\%). When asked what they hoped to achieve by sending appreciation (Figure~\ref{fig:kudos-achieve}), most senders (76\%) hoped to ``show that you value them as a person.'' Many users (38\%) hoped to offer feedback, while nearly a third saw themselves as paying an obligation or promising to reciprocate in the future.

\footnotetext{Margin of error on these responses in approximately 3-8\%.}

\begin{figure*}[ht!]
\centering
\subfigure[Messages sent per day]{
\includegraphics[height=1.35in]{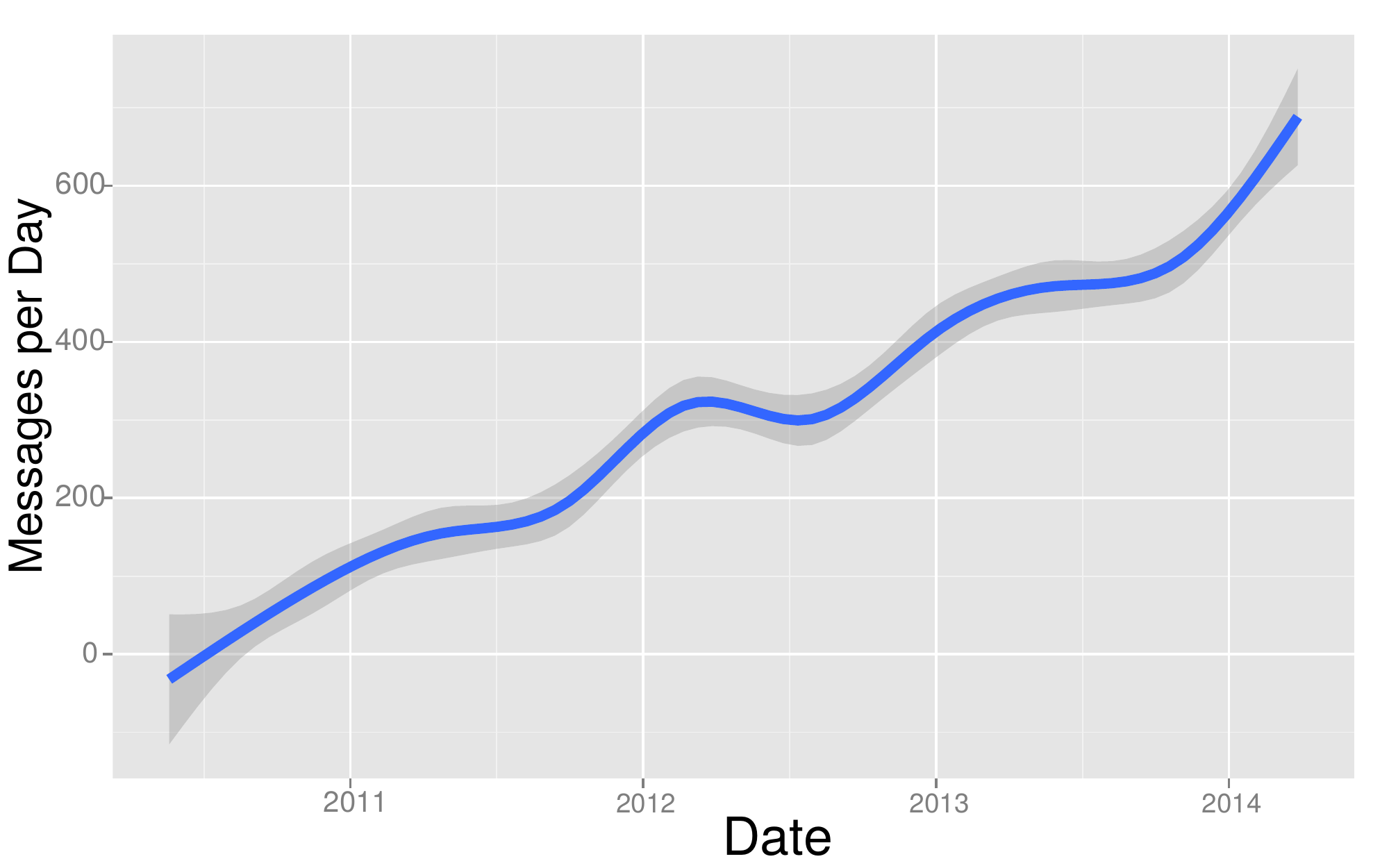}
}
\subfigure[Unique senders per day]{
\includegraphics[height=1.35in]{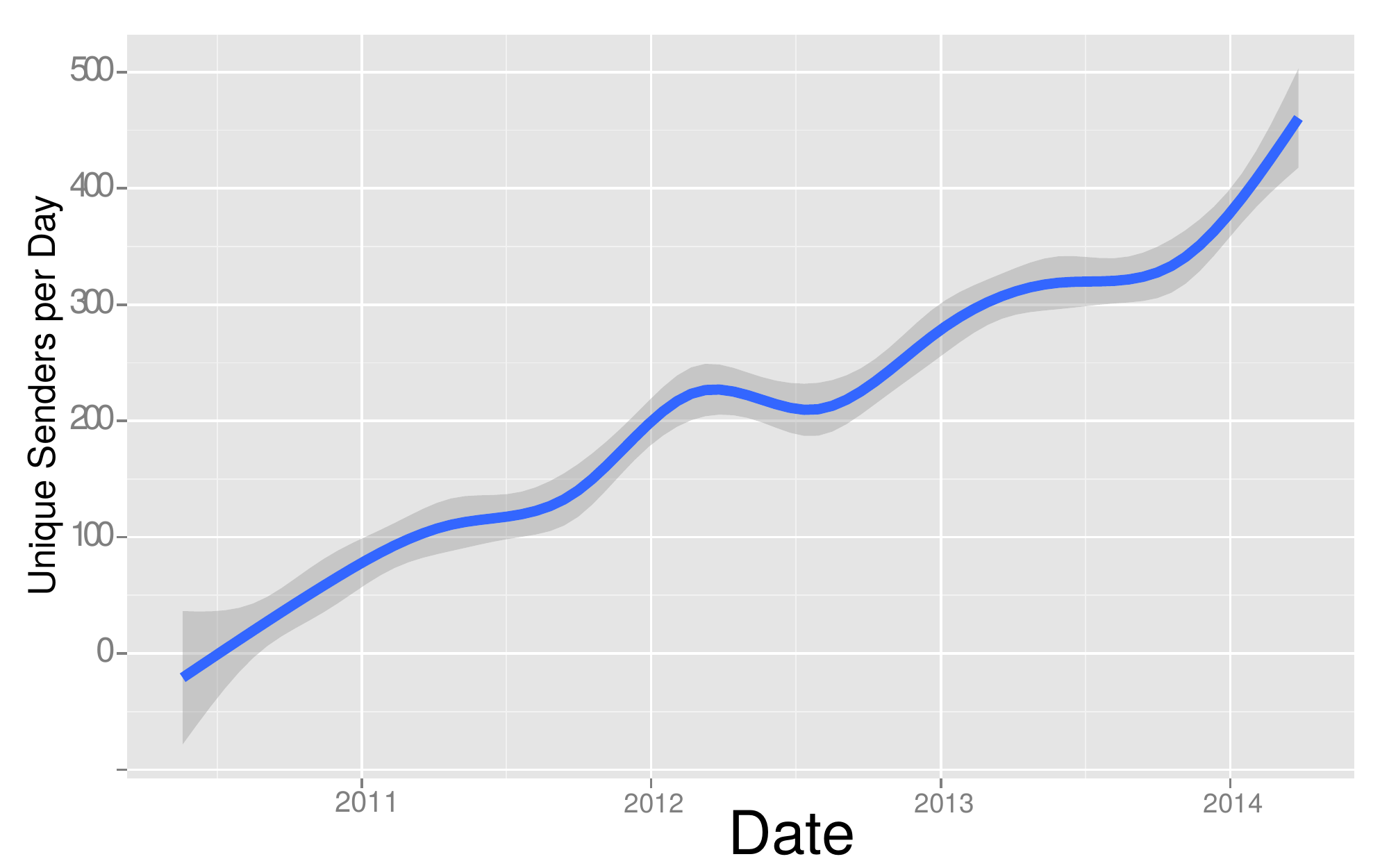}
}
\subfigure[Unique new and repeat senders per day]{
\includegraphics[height=1.35in]{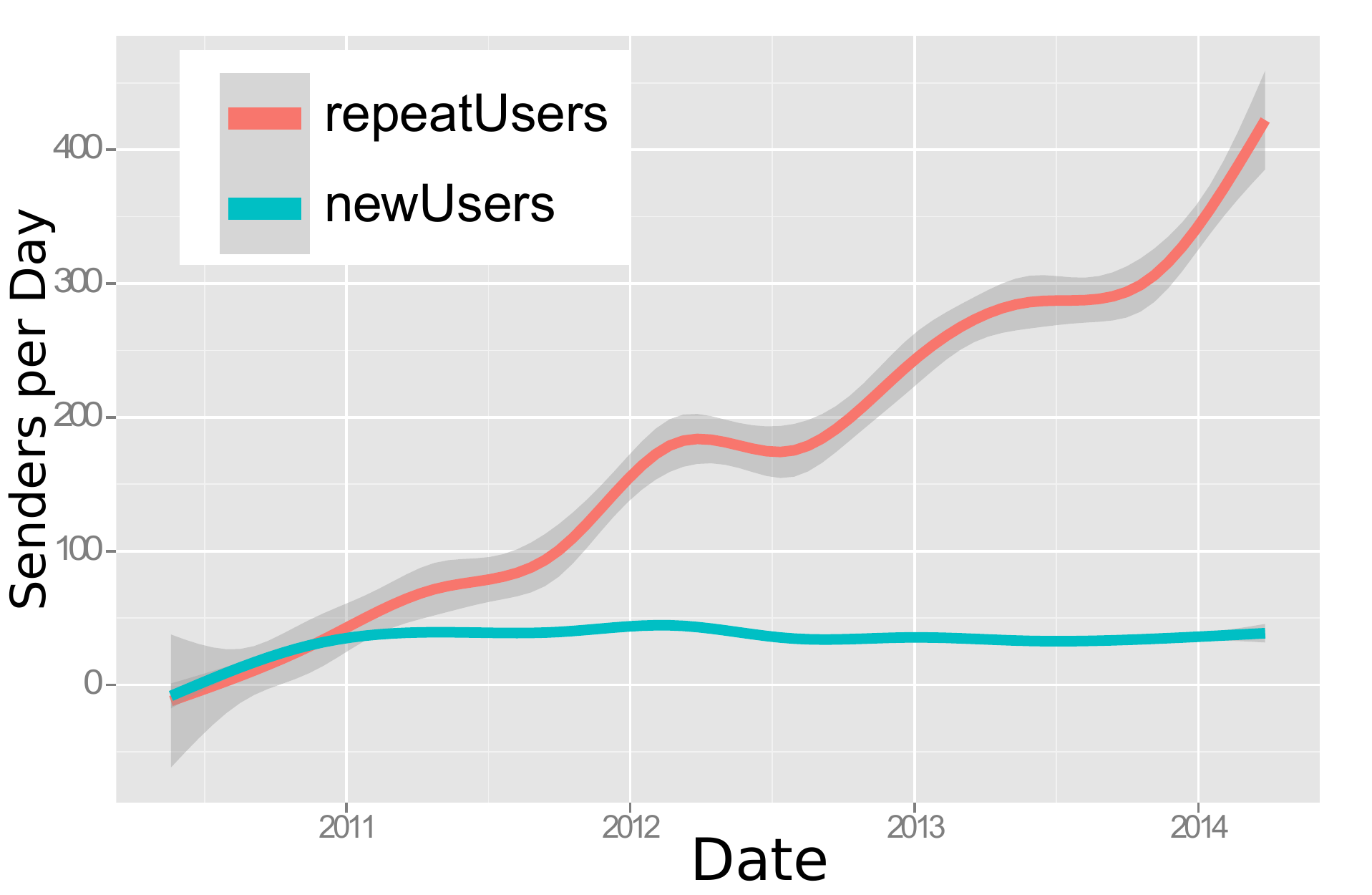}
}
\caption{Adoption and usage patterns. We consider messages sent per day, unique senders per day and unique new/repeat senders per day. Growth patterns are consistent across measures.}
\label{fig:usage}
\end{figure*}

\subsection{Descriptive Analysis of Adoption and Use}

Thanks messages were sent by over 47,000 unique senders to over 73,000 unique receivers. A total of just over 80,000 unique company employees had engaged with \emph{Gratia} at the time of our observation. Figure~\ref{fig:usage} shows (a) the average number of \emph{Gratia} messages sent per day, (b) the number of unique senders per day and (c) the number of new versus repeat senders per day over the period of observation. The data show increasing use and adoption over time. The pattern of messages sent and unique senders over time are steadily increasing across the entire observation period. In Figure~\ref{fig:usage}(c) we also see both consistent recruitment to the system and retention of users over time. New users join at a relatively constant rate, while the number of repeat users increases over time. Together these figures illustrate a clear pattern of increasing use, without any evident signs of leveling off. 

While senders of thanks messages primarily report (in survey data) learning about \emph{Gratia} via word of mouth (62\%), managers and receivers of thanks report learning about the system almost equally by word of mouth and receiving a \emph{Gratia} notification. 41\% of receivers report learning about it from a notification, and 39\% report word of mouth; 46\% of managers report learning about it by receiving a notification, the same percent as word of mouth. These word of mouth recommendations may have come from managers; 54\% of whom report encouraging their teams to use the system. Email signatures were another way users learned about the system; 8\% of managers, 11\% of receivers, and 14\% of senders reported learning about \emph{Gratia} from email signatures that linked to the system, like \emph{``Was my response helpful? \underline{Consider sending me a Gratia}.''} 

\subsection{Who Do Users Thank?}

Users have the opportunity to express appreciation via \emph{Gratia} to any of their co-workers, be they managers, peers or the company CEO. In practice, social norms are likely to structure thanks. We begin our investigation of thanks by considering the extent to which users thank back; reciprocity is a common social norm found many social systems. 

There are three potential relationship configurations that could exist between any pair of employees or \textit{dyad}: no relationship, a one-directional (or asymmetric) relationship, or a mutual relationship. 99.9\% of all \emph{Gratia} dyads have never interacted there therefore have no relationship. We also observe that \emph{there are twice as many asymmetric pairs as mutual pairs}, 0.006\% and 0.003\% of all pairs, respectively. 

To explore mutual relationships, we compare reciprocation against the baseline chance of a tie, taking the relative log-odds of a tie given a reciprocation, versus the baseline probability of a tie \cite{butts_social_2008}. This is a more appropriate measure of reciprocity in large, sparse networks when relationships have low probability.  We find that \emph{thanks messages are 9 times more likely to be sent between a pair in which a tie already exists, than between a random pair.} 

Despite strong reciprocity norms, 80.4\% of connected pairs in \emph{Gratia} only interact once (i.e. exchange one message). On the other extreme, one employee pair exchanges 68 thanks messages over the observation period. This pair are peers, both working under the same manager. For other employees who use \emph{Gratia} and thank each other repeatedly over time, only 5\% of pairs interact 3 or more times. In terms of the lifespan of relationships between pairs, the average time between first and last interaction is 231 days. On average individuals wait 124 days between messages, just over 4 months. When asked about the four-month interval between thanks in free-text questions, most senders were unsure. 17\% of senders argued that reciprocating thanks was not necessary. One sender explained that ``they aren't doing it just for reciprocation, but genuinely \textemdash and it may take time for a genuine need for returning \emph{Gratia} to occur.''

Aside from thanking someone back, how likely are users to exchange favors and other forms of support? Among senders, 31\% reported using \emph{Gratia} messages to confirm that they would be willing to help the receiver in the future (Figure~\ref{fig:kudos-achieve}). We asked senders to rate how likely they were to do a favor from someone they thanked on \emph{Gratia}, from ``as likely as a request from anyone" to ``only if I've sent them appreciation," and 65\% reported that they would be more likely to do a favor for someone they sent appreciation. 

\subsubsection{Appreciation Within and Across Teams}

Next, we investigate correspondence between the informal \emph{Gratia} social network and the formal, organizational network of managerial relationships. One of the primary motivations for analyzing these two representations of relationships in the workplace is to explore the structure of thanks expressions within and across teams. 

To do so, we consider two measures of the formal ``distance'' between the sender and receiver of a thanks message, each measured in the organizational hierarchy network. The first metric \textemdash which we call \emph{organization distance} \textemdash computes the length of the shortest path (a sequence of existing ties in the formal company network) between the two individuals. Since the organizational network is a strict tree, this involves tracing up the organizational chart from the initiating employee to a common authority and then down to the thanks recipient. The second metric \textemdash which we call \emph{hierarchy distance} \textemdash computes the difference in the sender's and receiver's level in the hierarchy; level is measured as steps below the top of the organizational tree, i.e. steps from the head of the company (again a step in this case is a managerial relationship).

To identify within-team and cross-team appreciation, we use the number of steps in the organizational hierarchy required to find a manager that has authority over both the sender and receiver of the \emph{Gratia} message. This is the \emph{organizational distance} between two employees, as described above. Two employees with an organizational distance of 2 would be coworkers, reporting to the same manager. Pairs with an organizational distance of greater than two would indicate coworkers in different teams. 

On average, the distance (path) between the sender and receiver of a \emph{Gratia} message is 7.7, meaning that thanks messages are rarely sent among members of the same direct team.  We find that only around 10\% of \emph{Gratia} messages are sent to fellow team members (a colleague with whom the sender shares a manager), reinforcing that thanks in \emph{Gratia} are typically send to coworkers outside of one's own team. The use of \emph{Gratia} across teams is consistent with survey results as well. In free-text replies, a commonly cited reason for sending messages (Figure \ref{fig:sender-reason}) is the physical distance between sender and receiver.

\subsubsection{Upstream, Downstream, and Peer Appreciation}

\begin{figure}[ht!]
\centering
\subfigure[Local \emph{Gratia} network]{\includegraphics[width=.45\linewidth]{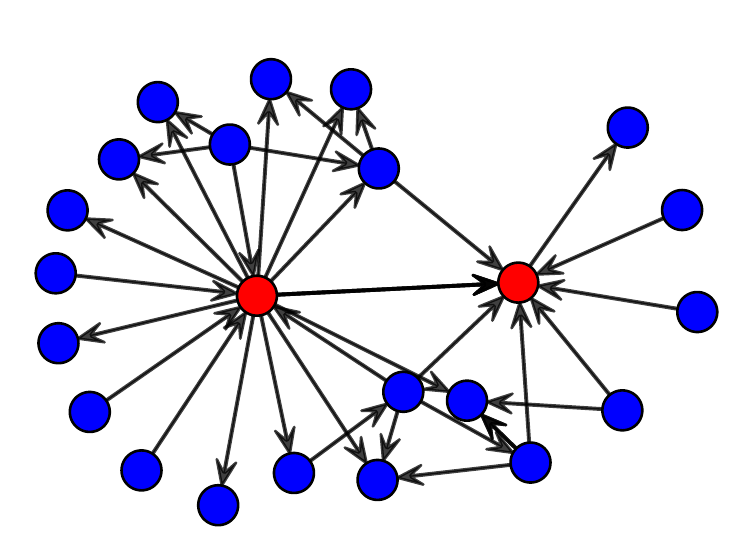}}
\subfigure[Formal hierarchy]{\includegraphics[width=.45\linewidth]{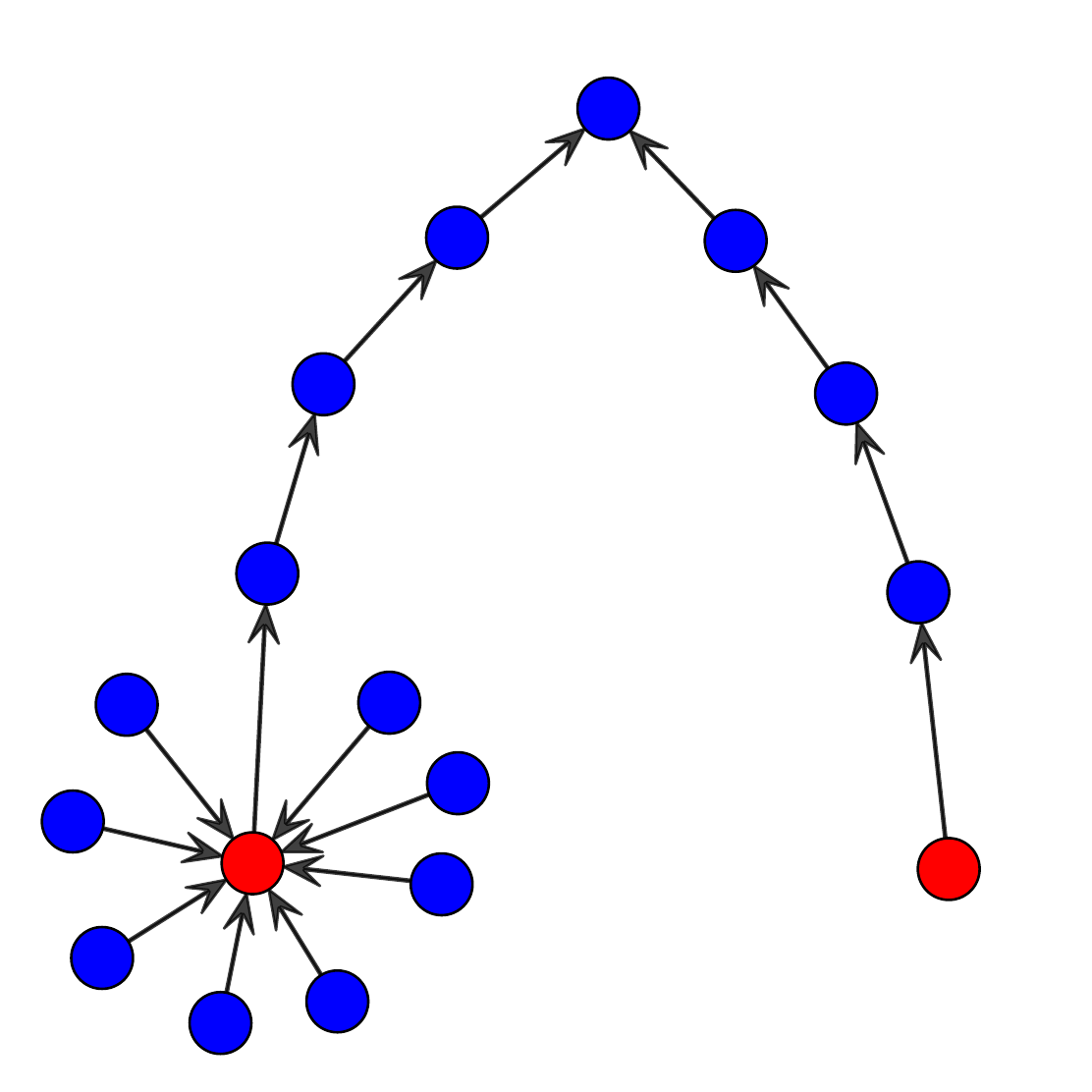}}
\caption{Local \emph{Gratia} network for illustrative thanks message, sent between the two red nodes, seen in (a). Other employees who have also sent or received a thanks message to/from these focal individuals are shown in blue. Path between the two focal individuals (red) in the formal organizational network shown in (b). Blue nodes represent all direct managerial relationships of focal individuals.}
\label{fig:example_distance}
\end{figure}

Are \emph{Gratia} messages more likely to be directed from manager to subordinate (downstream), from subordinate to manager (upstream), or between employees at similar levels? To investigate this question, we use the hierarchy distance metric, described earlier. The average difference in the hierarchy distance or corporate level between sender and receiver is 0.0, implying that \emph{Gratia} messages are primarily sent to peers of the same level. If more managers were sending appreciation to team members or their teams were sending messages to their managers, the distance would be closer to 1.0. We illustrate this result, as well as the previous on organizational distance, in Figure~\ref{fig:example_distance}, showing the local, informal \emph{Gratia} network and corresponding organizational network for two focal participants. In combination these results could suggest that appreciation systems like \emph{Gratia} help to facilitate an expression of thanks when distance (physical or social) separates individuals who are at similar levels in the company. 

When senders were asked in free-text questions why upstream appreciation was uncommon, most respondents were unsure, although 21\% of senders cited possible negative outcomes for public, upstream appreciation. For example, one respondent wrote that ``\emph{Gratia} is public and it might be seen as `asskissing' :)." Likewise, other respondents mentioned this might be perceived as ``brown-nosing.' Senders were mostly unsure why downstream appreciation was uncommon, but we can presume that managers might not send messages in a system whose primary feature involved resending copies of their own appreciation back to them.

\subsection{What Outcomes do Users Expect from Participation?}

\begin{figure}[ht!]
\includegraphics[width=\linewidth]{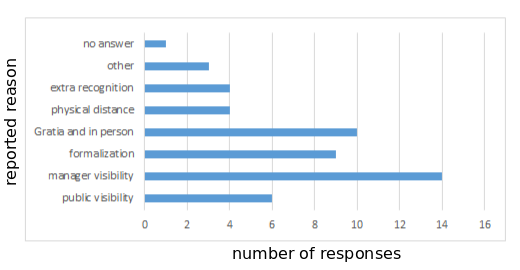}
\caption{User survey responses when asked about reasons to use \emph{Gratia} rather than thank in-person.}
\label{fig:sender-reason}
\end{figure}

Users report a variety of reasons for expressing thanks via the \emph{Gratia} system. User-reported motivations for use, seen in Figure \ref{fig:sender-reason}, often revolve around visibility and formalization. The primary reported reason for using the system was that the recipient's manager would also be notified of thanks.\footnote{Open-ended survey questions were categorized by two independent coders after a mutually agreed upon coding scheme was developed. The coding scheme was developed via an iterative, grounded approach. Categories were determined by examining responses and expanded when necessary. Each codes then applied the codes to the entire set of responses independently. Inter-rater reliability measure shows higher level of agreement, with percent agreement over 89\% for all codes and Cohen's Kappa values for each code of at least 0.60 (majority over 0.80).}

Not everyone, however, realized what managers could see. In response to multiple choice questions, 87\% of receivers agreed that their immediate managers could see the messages they received, but 52\% of senders were unsure if their managers could see the messages they sent. This uncertainty among senders about who sees their messages is reflected in their expectations of the personal outcomes of sending appreciation. Among senders, only 14\% expected that sending appreciation would benefit their performance review, and 76\% expected that it would not do so. More receivers believed that senders would gain benefits too; 30\% of receivers expected that senders would experience benefits to their performance reviews, with only 46\% disagreeing. 

Do managers actually see these messages? Managers whose team members send or receive appreciation tend to encounter \emph{Gratia} messages through email notifications. Among managers, 54\% report reading the email notifications, but only 8\% report visiting the system website. Less than half of managers report being aware of exactly how many \emph{Gratia} their team members receive.  

\subsubsection{The Role of Appreciation in Manager Decisions}

Managers use appreciation information to support employee feedback, promotion decisions, team formation, and performance reviews. 61\% of managers who received notifications report incorporating appreciation into decisions affecting their team members. Out of these, the most common use was offering feedback to team members (54\%). A small minority of managers report incorporating appreciation into decisions about team formation (8\%) and promotion or restructuring decisions (8\%). 39\% of managers report not using this information for any decisions. Sender and receiver's expectations of direct performance review benefits may have been overly optimistic. While 57\% of receivers expected that their performance view would be affected and 62\% of senders expected the same, only 15\% of managers report incorporating the information in performance reviews. \emph{Gratia} did bring other, reputational benefits to senders and receivers: many managers (77\%) do report discussing appreciation received by their team members. 31\% have mentioned it to the receiver, 23\% have mentioned appreciation to a third party, and 23\% have mentioned a \emph{Gratia} message to their entire team.

\begin{figure}[ht!]
\centering
\subfigure[Daily seasonality]{
\includegraphics[width=.7\linewidth]{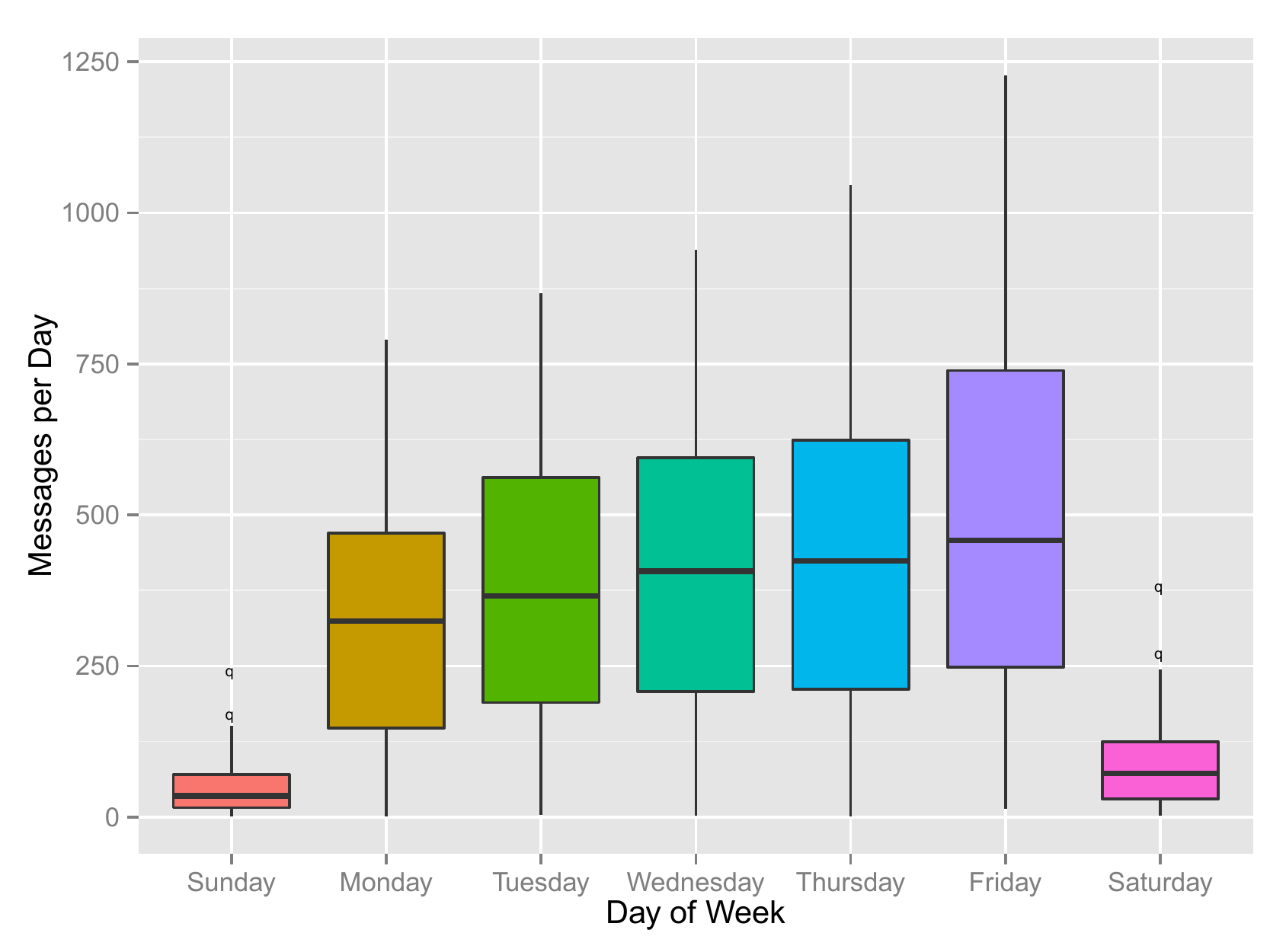}}
\subfigure[Monthly seasonality]{
\includegraphics[width=.7\linewidth]{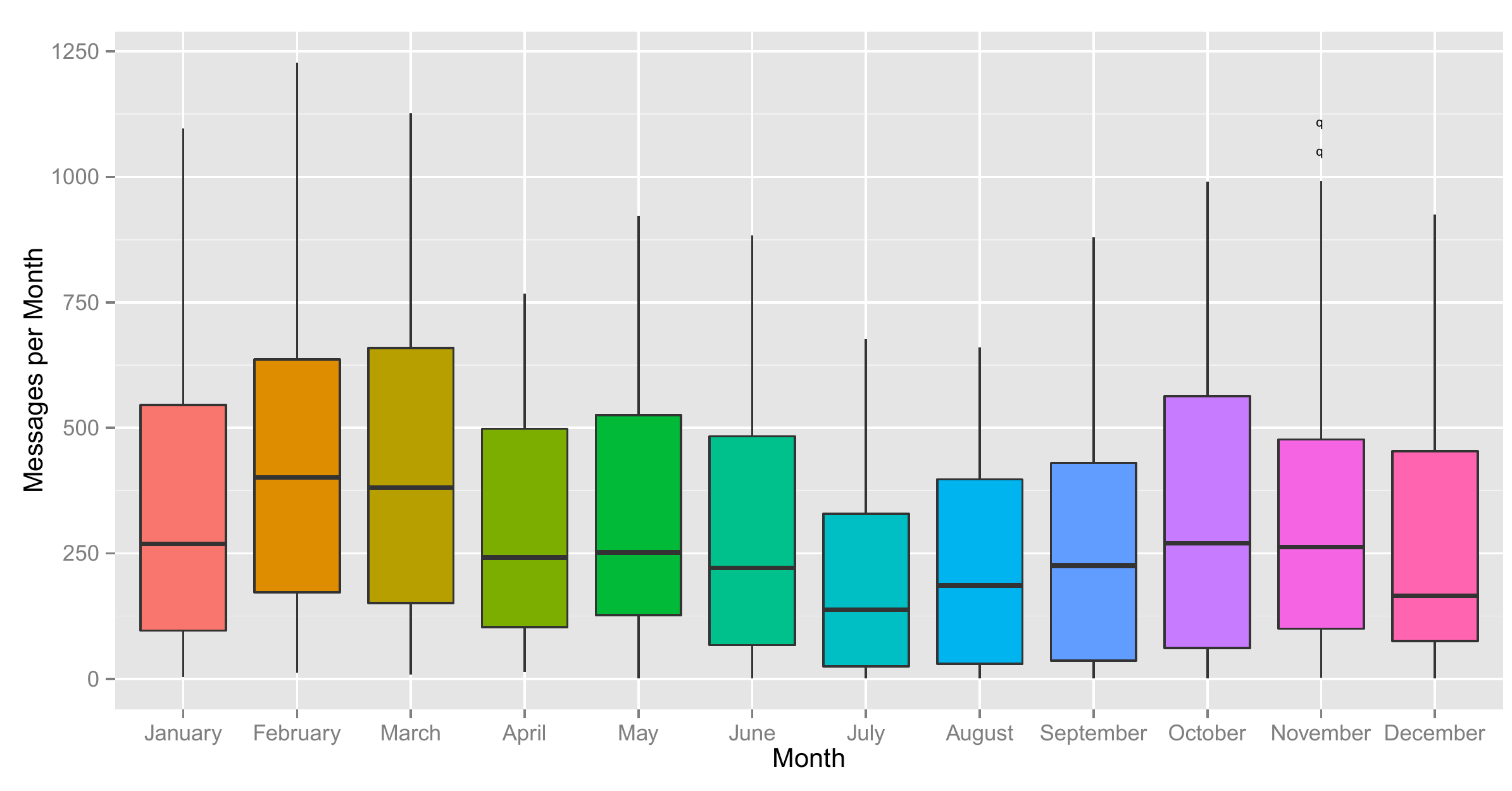}}
\caption{General usage and seasonality in messages sent over time.}
\label{fig:seasonality}
\end{figure}

User beliefs about the influence of \emph{Gratia} on manager decisions is reflected in system usage patterns. As with most communication behaviors, expressions of thanks demonstrate seasonality. Figure~\ref{fig:seasonality}(a) shows the distribution of messages sent by day of week; use is characterized by an increase in activity at the end of the work week, Friday in particular shows high rates of thanks sent. As one might expect, expression of thanks drop during weekends. Figure~\ref{fig:seasonality}(b) shows the distribution of messages sent by month of the year. Fewer messages are sent during summer months. Highest rates of messages sent occur in February and March, coinciding with the period before employees' performance appraisals. When asked about these monthly patterns in free-text questions, most senders had no explanation. However, 24\% reported that more \emph{Gratia} were likely sent in advance of mid-year performance reviews.

\subsubsection{Rewards and Bonuses}

We asked senders ``what are your thoughts about using a system like \emph{Gratia} to influence rewards or bonuses?" and coded responses for positive, negative, or no answer. Senders were overall in favor of using a system like Gratia to give rewards, with 59\% in favor and 31\% opposed. One person even suggested a specific dollar amount: ``\$10 per 1 Gratia received, \$5 per 1 Gratia sent paid monthly." However, most responses came with caveats, such as ``should be considered but not the only factor." The senders who were not in favor of linking \emph{Gratia} to compensation feared abuse, and losing the wholesomeness of the system: ``if you add money to the mix, you will encourage gaming. There is something refreshingly pure about the Gratia system \textemdash whether getting or giving, the currency is pure appreciation, which is rare.''

\section{Discussion}
In this paper, we identify the growing genre of appreciation systems, workplace messaging and microblogging systems that mediate and track expressions of appreciation. We describe and compare the design affordances of 13 of these systems. In addition, we conduct a case study of survey responses and user logs of 422,000 appreciation messages sent within the \textit{Gratia} system over four years.

Appreciation systems are in part designed to address a wide problem of recognition. These systems can only address that problem if the thanks on those systems is exchanged between workers across teams, where managers can less easily observe their work. In our case study of \textit{Gratia}, we ask if these systems follow hierarchical feedback structures, ``democratize kissing up," or reach across teams. We find that thanks messages are typically sent to coworkers outside of one's own team, and that they are sent across the company hierarchy rather than up or down it.

Although usage of \textit{Gratia} satisfies the structural requirements of a response to the recognition problem, our survey results suggest that it may face its own recognition problem. Users report manager visibility as the main reason for using \textit{Gratia}, and most managers do report seeing the messages and mentioning them to team members. However, while users had high expectations for the effect of their messages on manager decisions and even sent more appreciation close to the performance review period, a large minority of managers ignored appreciation reports and did not use them to inform decisions. The majority of managers, who did incorporate appreciation into their decisions, mentioned \textit{Gratia} in feedback or conversations rather than performance reviews or promotion decisions.

\section{Conclusion and Future Work}

This paper offers an overview of the appreciation system genre and focuses on quantitative analysis of logs and surveys for a single system. We have described the structure of thanks and user expectations in a case study of \emph{Gratia}, a system used within a large multinational corporation. The observed high, and growing, use of the \emph{Gratia} system may represent a gap in the social affordances of messaging and SNS systems already used in the company. Thanks systems may be especially useful to mediate gratitude among distributed teams working on large projects, for whom in-person thanks might be less possible. Care should be taken however, when incorporating thanks messages into employee reviews, to account for the patterns of thanking in relation to the organization chart, and the lack of thanks between direct peers.

Future research could attempt to replicate our findings across multiple companies and different appreciation systems. Other studies could if sending or receiving thanks have an effect on well-being, productivity, chances of promotion, or employee retention. Other research could compare the effects of reward and badge systems to those that do not include rewards. Qualitative research with workers and managers may yield contributions to theories of relation work and offer further nuance to the choices, rationales, and trade-offs associated with the affordances we have identified. Finally, for corporate SNS that have added appreciation features after initial deployment, researchers may find natural experiments for estimating the effect of peer appreciation on the recognition problem in the adoption of these systems in the workplace.

\small
\bibliographystyle{aaai}
\bibliography{gratia_icwsm}
\end{document}